\documentclass[a4paper,12pt,fleqn]{scrartcl}
\usepackage{graphicx}
\usepackage{amsmath}
\usepackage{amssymb}
\usepackage{bm}
\usepackage{fourier}
\setkomafont{sectioning}{\fontencoding{OT1}\fontfamily{phv}\fontshape{sf}\fontseries{b}\selectfont}
\usepackage{wrapfig}

\begin{document}

\section*{In memoriam: Markus B\"{u}ttiker (1950 -- 2013)}

\begin{wrapfigure}{r}{0.5\textwidth}
  \begin{center}
    \includegraphics[width=0.48\textwidth]{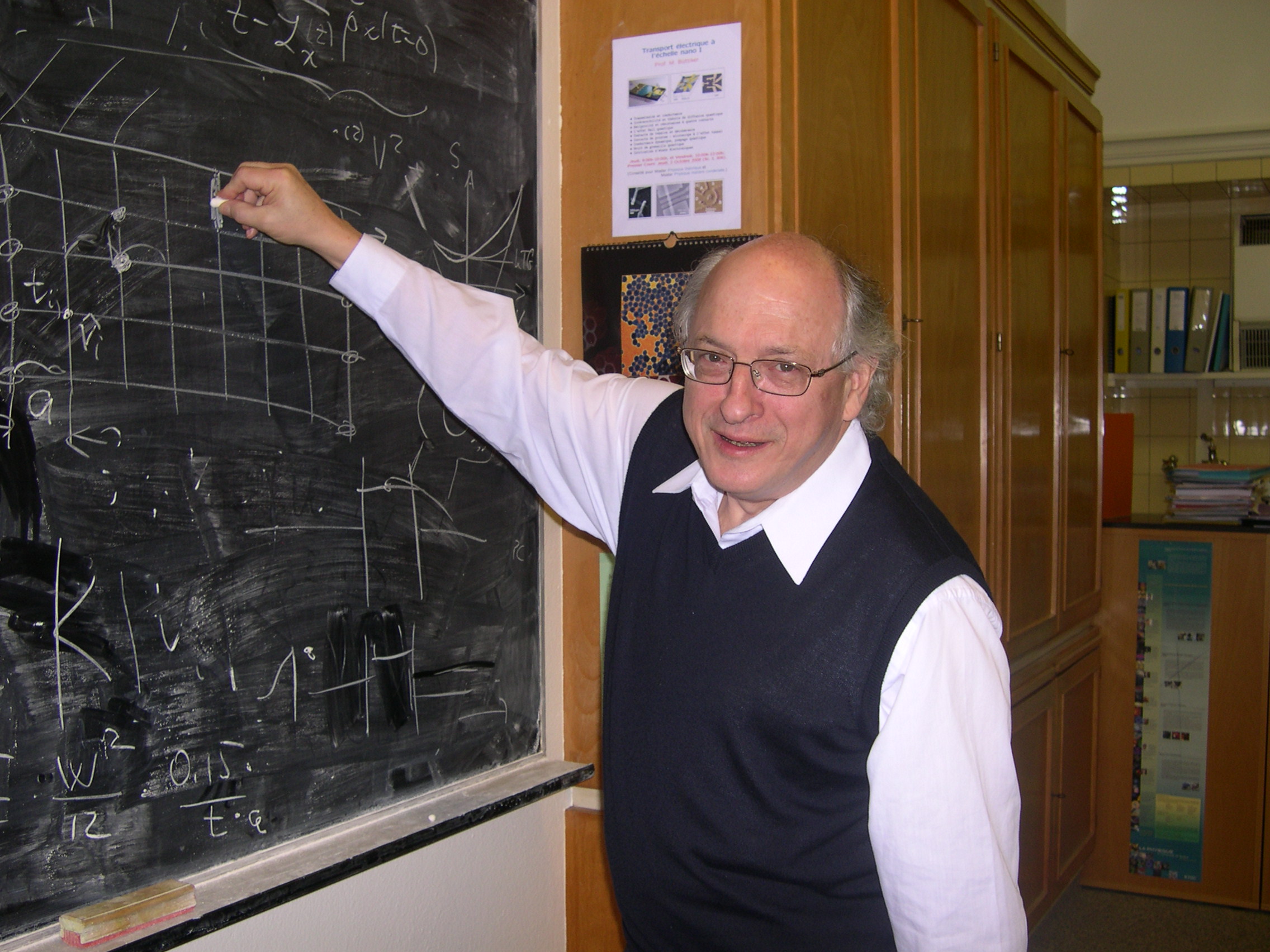}
   \textit{\small Markus B\"{u}ttiker in his office in Geneva,\\
   January 2009. [Picture taken by Teresa Dib.]}
  \end{center}
\end{wrapfigure}

This issue\footnote{\textit{ Special issue dedicated to the memory of Markus B\"{u}ttiker}, Physica E \textbf{82}, 1--374 (2016).} of Physica E collects contributions from the frontier of quantum electronic transport, a field that owes so much to the pioneering work of Markus \mbox{B\"{u}ttiker}. The authors include former students, postdocs, and academic collaborators of Markus, a sampling from the community of mesoscopic physicists that counts him among its founders. 

It is in the nature of the most influential research to become assimilated into common knowledge, to become obvious. The scattering formulation of electrical conduction seems so natural and self-evident today, it is difficult to appreciate the controversy and confusion that surrounded its emergence in the 1980's. Experiments on small and weakly disordered conductors could no longer be explained in terms of a local relation between current and electric field. The very nature of ``electrical resistance'' was debated, how can one associate a resistance to a perfect conductor? Experiments showed that the voltage measurement itself had become a significant perturbation, but was it really necessary to include the voltmeter into the theory? The first quantum interference effects were observed, and it appeared that the well-established Onsager symmetry relations for conductivity tensors were violated. Or were they just mis-interpreted?

The complete and convincing resolution of these issues came from a small group of researchers at IBM, centered around Rolf Landauer and his junior collaborator Markus \mbox{B\"{u}ttiker}, with Yoseph Imry as a regular visitor. The theoretical framework is now referred to as the ``Landauer-B\"{u}ttiker formalism'', a name Markus disliked because it drew attention to the formal aspects (linear algebra of scattering matrices), rather than to the underlying physical concepts. His intuition that a voltmeter is ``an electron reservoir that draws no current from an ideal lead connecting it to the conductor'' seems unremarkable today, but it was a radical departure from the prevailing notion that a voltmeter ``measures an integral of the electric field between two points''. By treating current sources and sinks on equal footing with voltage probes, Markus arrived at a nonlocal description of electrical conduction that was fully consistent both with experiments and with the Onsager symmetry relations \cite{1}.

The power of this approach was demonstrated dramatically in the context of the quantum Hall effect \cite{2}. Markus was characteristically uncomfortable with Laughlin's profound explanation of the exact quantization of the Hall conductance. Surely a complete theory must be able to describe both the ideal and the non-ideal situation, and indeed experiments on mesoscopic Hall bars did find deviations from exact quantization. B\"{u}ttiker's voltage-probe model could explain the observed deviations in terms of absence of local equilibrium at the edge of the conductor. A debate ensued on whether the quantum Hall effect was a bulk effect or an edge effect. Ever practical, Markus would steer the debate from semantics to observable consequences of edge state transport (leaving phrases as ``bulk-edge correspondence'' to more formally inclined theorists).

These theoretical developments went hand-in-hand with experimental progress, but another topic from the IBM period was further ahead of experiments: The prediction with Imry and Landauer of a persistent current in a normal-metal ring \cite{3}. The calculation for an ideal impurity-free ring was elementary, the fundamental new insight was that elastic scattering by disorder would not lead to a decay in time of the circulating current. That elastic scattering is not a source of dissipation is again an example of a notion that has been assimilated and seems entirely obvious today --- but was originally counter-intuitive.

Time-dependent properties of mesoscopic conductors would be a unifying theme of much of the research of Markus B\"{u}ttiker after he left the IBM Research Laboratory for a faculty position at the University of Geneva. The scattering approach was extended from the time-averaged current to current fluctuations, resulting in the prediction of the 1/3 suppression of shot noise \cite{4} and summarized in an influential review \cite{5}. He developed a dynamical theory of mesocopic capacitors \cite{6,7} and periodically driven quantum pumps \cite{8}. 

With the rising interest in quantum information processing, Markus started to apply the scattering approach to entangled electrons \cite{9}. He had many exciting ideas in this direction, which he proposed in the project ``Floquet Computers'' for a major European research grant. The news that this grant was awarded arrived when he was already hospitalized. We now have to miss his insight, creativity, and friendship.\medskip\\

\noindent
\textit{The guest editors of this special issue:\\
Carlo Beenakker, Philippe Jacquod, Andrew Jordan, and Peter Samuelsson.}

\end{document}